\documentclass[]{aa} 

\usepackage{graphicx}
\usepackage{txfonts}
\begin{document}
   \title{The geometry of the close environment of SV\,Psc as probed by VLTI/MIDI\thanks{Based on observations made with ESO telescopes at La Silla Paranal Observatory under program IDs 082.D-0389 and 086.D-0069.}}


   \author{D. Klotz\inst{1}, S. Sacuto\inst{2}, F. Kerschbaum\inst{1}, C. Paladini\inst{1}, H. Olofsson\inst{3}, \and J. Hron\inst{1}
          }

   \institute{Department of Astrophysics, University of Vienna,
              T\"urkenschanzstrasse 17, A-1180 Vienna\\
              \email{daniela.klotz@univie.ac.at}
         \and
             Department of Physics and Astronomy, Division of Astronomy and Space Physics, Uppsala University, Box 516, 75120, Uppsala, Sweden
             \and
             Onsala Space Observatory, Dept. of Earth and Space Sciences, Chalmers University of Technology, 43992, Onsala, Sweden
             }

   \date{Received ; accepted}

 
  \abstract
   {SV\,Psc is an asymptotic giant branch (AGB) star surrounded by an oxygen-rich dust envelope. The mm-CO line profile of the object's outflow shows a clear double-component structure. Because of the high angular resolution, mid-IR interferometry may give strong constraints on the origin of this composite profile.}
   {The aim of this work is to investigate the morphology of the environment around SV\,Psc using high-angular resolution interferometry observations in the mid-IR with the Very Large Telescope MID-infrared Interferometric instrument (VLTI/MIDI).}
   {Interferometric data in the $N$-band taken at different baseline lengths (ranging from 32--64\,m) and position angles (73--142$^\circ$) allow a study of the morphology of the circumstellar environment close to the star. The data are interpreted on the basis of 2-dimensional, chromatic geometrical models using the fitting software tool GEM-FIND developed for this purpose.}
   {The results favor two scenarios: (i) the presence of a highly inclined, optically thin, dusty disk surrounding the central star; (ii) the presence of an unresolved binary companion at a separation of $13.7^{+4.2}_{-4.8}$\,AU and a position angle of 121.8$^\circ$$^{+15.4^\circ}_{-24.5^\circ}$ NE. The derived orbital period of the binary is $38.1^{+20.4}_{-22.6}$\,yr. This detection is in good agreement with hydrodynamic simulations showing that a close companion could be responsible for the entrainment of the gas and dust into a circumbinary structure. }
   {}

    \authorrunning{Klotz et al.}          
    \titlerunning{The circumstellar environment of SV\,Psc}

   \maketitle
%

\section{Introduction}
Stars on the asymptotic giant branch (AGB) undergo phases of heavy stellar mass loss, which lead to the formation of circumstellar envelopes. Early results obtained from interferometric maps of OH maser emission (e.g. Booth et al. 1981; Bowers \& Johnston 1990) as well as CO radio line maps (Truong-Bach et al. 1991) were pointing to an overall spherically symmetric mass-loss. However, the shape of a significant fraction of the CO line profiles of AGB stars deviates significantly from that expected of a spherical envelope (e.g. Knapp et al. 1998; Olofsson et al. 2002; Winters et al. 2003). Such a deviation is evident in the case of the oxygen-rich star SV Psc, which shows one of the more extreme CO line shapes (Kerschbaum \& Olofsson 1999; Winters et al.\,2003). \\
SV Psc is a semiregular variable star (SRb) located at a distance of $\sim$400\,pc (Olofsson et al. 2002; Winters et al. 2003). For the distance determination the authors assumed a luminosity of the star of 4\,000 and 6\,000\,$L_{\odot}$, respectively. The mid-infrared dust emission features in the ISO/SWS data (Sloan et al.\,2003) indicate an oxygen-rich, dusty environment. Kerschbaum \& Olofsson (1999) and Olofsson et al. (2002) found that the CO $J$\,=\,1--0 and $J$\,=\,2--1 lines show a clear double component in the outflow emission, where a very narrow feature ($\sim$\,1.5\,km\,s$^{-1}$ for $J$\,=\,1--0, $\sim$\,2.3\,km\,s$^{-1}$ for $J$\,=\,2--1) is centered on a much broader one ($\sim$\,9.5\,km\,s$^{-1}$for $J$\,=\,1--0, $\sim$10.8km\,s$^{-1}$ for $J$\,=\,2--1). This two-component feature was confirmed by Winters et al. (2003), but they reported slightly higher expansion velocities and mass-loss rates of $1.2\times 10^{-7}$\,$M_{\odot}$\,yr$^{-1}$ (narrow component) and $1.1\times10^{-6}$\,$M_{\odot}$\,yr$^{-1}$ (broad component). SV\,Psc was also detected in the SiO radio line emission observations ($J$\,=\,2--1, 5--4) by Gonzalez Delgado et al.\,(2003), where the two-component structure is visible again.\\
Many scenarios have been proposed to explain this kind of line profile:
(i) Knapp et al.\,(1998) interpret it as two consecutive winds with different age, expansion velocity and mass-loss rate. The narrow feature corresponds to a more recent mass-loss episode in this case. Winters et al. (2007) noted that this kind of two-wind structure may apply also to the O-rich AGB star EP\,Aqr, which shows CO line profiles similar to those of SV\,Psc. 
(ii) Bergman et al. (2000) found that the two-component CO line profile around the O-rich AGB star RV\,Boo can be caused by a low-mass disk in Keplerian rotation. They interpret the narrow feature as a spherically symmetric wind close to the star. The presence of a disk around this star was confirmed by Biller et al. (2005) using mid-IR adaptive optics images. Jura \& Kahane (1999) argued that the narrow CO line profiles found in a number of red giant stars could be caused by long-lived reservoirs of orbiting molecular gas (note that these stars do not show double-component profiles). They interpret the disk as being caused by matter from previous mass-loss events that is captured by a binary system. This would mean that the disk is a direct consequence of a close binary component. 
The formation of a circumbinary disk was also predicted by three-dimensional hydrostatical models of Mastrodemos \& Morris (1999). However, Knapp et al.\,(1998) claim that the frequency of double winds appears to depend on the stellar chemistry and variable type and is therefore unlikely to be the result of binarity. 
(iii) Kahane \& Jura (1996) suggested a very slow, spherically expanding wind in combination with a bipolar outflow for the AGB star X\,Her, where the latter would be causing the broad plateau emission. Nakashima (2005) was able to confirm that the broad component in the CO line profile could indeed be caused by a bipolar outflow and the narrow component is caused by a rotating disk (from interferometric observations of the CO $J$\,=\,1--0 line). Castro-Carrizo et al. (2010) also found evidence for the bipolar outflow in the $^{12}$CO line emission, but did not detect the Keplerian disk that was suggested by Nakashima (2005).\\
The origin of these composite profiles, even though extensively studied, is still puzzling. High-angular-resolution observations in the mid-IR provide the possibility to study the morphology of the close circumstellar environment, hence providing constraints on the mechanism responsible for such a line profile. In this paper, we present the first mid-IR spectro-interferometric VLTI/MIDI observations of the AGB star SV\,Psc, and attempt to find the most probable among the proposed scenarios. \\
\\
In Sect.~\ref{observations} an overview of the observations of SV\,Psc as well as on the data reduction is given. The geometrical models and their interpretation are presented in Sect.~\ref{models}. A discussion of the results and the conclusions are given in Sect.~\ref{discussion} and \ref{conclusion}.


\section{Observations}
\label{observations}
\begin{table*}[htdp]
\caption{\label{journal}Journal of the MIDI Auxiliary Telescopes observations of SV\,Psc. }
\centering
\vspace{0.5cm}
\begin{tabular}{llllllllll}
\hline
\hline
Target & UT date \& time & Phase [cycle]  & Config. & $B_p$ & PA & Seeing & Mode \\
& & && [m] & [$^\circ$]&['']\\
\hline
SV Psc&2008 Nov 03 02:17:44&0.63 [-1]&H0-D0&54.5&81&0.44&HIGH-SENS\\
YY Psc&2008 Nov 03 03:25:53&\ldots&\ldots&\ldots&\ldots&0.93&SCI-PHOT\\
\hline
SV Psc&2008 Nov 03 04:14:19&0.63 [-1]&H0-D0&64.0&73&1.3&HIGH-SENS\\
YY Psc &2008 Nov 03 03:25:53&\ldots&\ldots&\ldots&\ldots&0.93&SCI-PHOT\\
\hline
SV Psc\tablefootmark{a} &2008 Oct 04 07:17:22&0.38 [-1]&K0-G1&82.7&38&1.46&HIGH-SENS\\
\hline
SV Psc\tablefootmark{a} &2008 Oct 04 08:18:23&0.38 [-1]&K0-G1&87.3&37&2.07&HIGH-SENS\\
\hline
SV Psc &2008 Dec 06 01:04:13&0.91 [-1]&D0-G1&65.0&126&0.75&HIGH-SENS\\
HD 12929&2008 Dec 06 00:11:09&\ldots&\ldots&\ldots&\ldots& 0.76&\ldots\\
\hline
SV Psc &2008 Dec 06 01:50:58&0.91 [-1]&D0-G1&59.7&125&0.59&HIGH-SENS\\
HD 12929&2008 Dec 06 00:11:09&\ldots&\ldots&\ldots&\ldots& 0.76&\ldots\\
\hline
SV Psc&2010 Oct 09 05:34:25&0.69 [5]&H0-I1&31.6&142&0.62&HIGH-SENS\\
$\alpha$\,Cet&2010 Oct 09 05:14:34&\ldots&\ldots&\ldots&\ldots&0.69&\ldots\\
$\alpha$\,Cet&2010 Oct 09 05:59:17&\ldots&\ldots&\ldots&\ldots&0.60&\ldots\\
\hline
SV Psc\tablefootmark{b}&2010 Dec 08 01:57:46&0.21 [6]&H0-I1&30.4&144&0.86&HIGH-SENS\\
\hline
SV Psc&2011 Jan 05 01:56:59&0.45 [6]&I1-E0&48.2&92&1.02&HIGH-SENS\\
$\alpha$\,Cet&2011 Jan 05 01:40:14&\ldots&\ldots&\ldots&\ldots&0.94&\ldots\\
$\alpha$\,Cet&2011 Jan 05 02:14:01&\ldots&\ldots&\ldots&\ldots&0.75&\ldots\\
\hline
\end{tabular}
\tablefoot{
Calibrators that were used to calibrate the data are given below the science target. The visual phase is given.  The baseline configuration, projected baseline length and position angle of the observations are given. The observatory seeing gives the average atmospheric conditions present at the observatory during the observation.\\
\tablefoottext{a,b}{Data not used (see Sect.~\ref{observations}).}}
\end{table*}

\begin{figure}
\centering
\includegraphics[width=8.cm,bb=112 382 458 709]{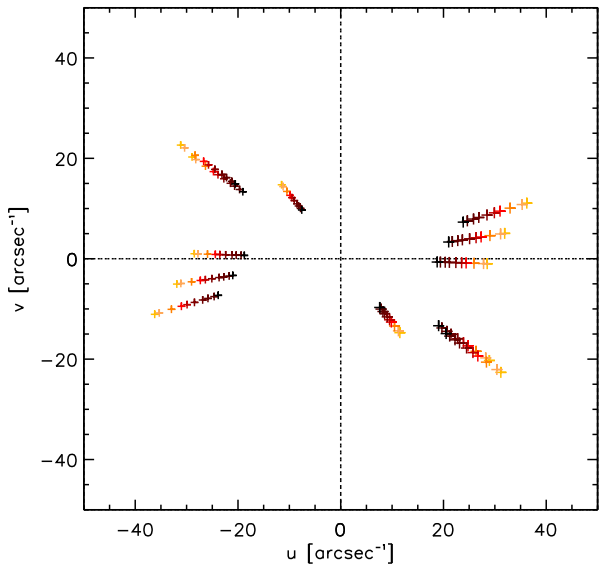}
\caption{\label{uv-coverage}$N$-band spectrally dispersed $uv$-coverage of the MIDI observations of SV\,Psc (see Table~\ref{journal}). Contour levels range from 8-12.5\,$\mu$m (light to dark, respectively) with a step size of 0.5\,$\mu$m.}
\end{figure}

\object{SV Psc} was observed in 2008 (Prog.\,082.D-0389) and 2010/2011 (Prog.\,086.D-0069) with the 1.8\,m Auxiliary Telescopes of the Very Large Telescope Interferometer MIDI. MIDI provides spectrally resolved visibilities, photometry and differential phases in the $N$-band (8--13\,$\mu$m). The baselines used range from 30--90\,m with position angles ranging from 36--144\,$^{\circ}$. The different baseline lengths provide the possibility to study the dusty envelope at different scales. The angular coverage (see Fig.~\ref{uv-coverage}), on the other hand, gives information on the morphology of the object.\\
The journal of available MIDI observations is given in Table~\ref{journal}. All science and calibrator observations (with the exception of the calibrator YY\,Psc that was observed in SCI-PHOT mode) have 
been carried out in HIGH-SENS mode and in low spectral resolution ($R=30$).  \\
In Sect.\,\ref{datared} the data reduction procedure and the determination of errors are described. A detailed discussion of data quality and different selection criteria is given in Sect.\,\ref{quality}. The $N$-band spectrometric variability of SV\,Psc is discussed in Sect.\,\ref{nbandvar}. 

\subsection{Data reduction}
\label{datared}
Data were reduced using the software packages\,\footnote{\tt{http://www.strw.leidenuniv.nl/$\sim$jaffe/ews/MIA+EWS-\\Manual/index.html}} MIA+EWS 1.7.1 (Jaffe 2004, Ratzka 2005, Leinert 2004). The methods used correspond to a coherent (EWS) and incoherent (MIA) analysis. The coherent analysis implements an off-line fringe tracking algorithm that aligns the interferograms before co-adding them. This results in a better signal-to-noise ratio of the visibility amplitudes. This method also provides differential phases. In the incoherent method the power spectral density function of each scan is integrated, resulting in a squared visibility amplitude. The squared visibility amplitude is then integrated over time. The differences in the output of MIA and EWS should always be within 5-10\% (Chesneau 2007). If this is not the case, according to the standard procedure, it is necessary to check if one of the masks used to convert the 2-dimensional frames to 1-dimensional spectra is not suitable for the data reduction. EWS uses a fixed mask, whereas MIA creates an adaptive mask. In this work, data were first reduced with MIA (with MIA and EWS mask), and afterwards with EWS (with MIA and EWS mask), and finally compared with each other.\\
The uniform-disk angular diameters used to derive the theoretical visibility for the calibrators as well as the corresponding IRAS 12\,$\mu$m flux and the spectral type are given in Table~\ref{calibrators}.\\
The final calibrated visibilities are derived by averaging the MIDI visibility measurements obtained with calibrators close to the science target observing time. Using measurements taken close in time ensures that the atmospheric conditions are comparable for the science and calibrator star. In this way, the transfer function should remove not only most of the instrumental errors, but also most of the errors introduced by the Earth's atmosphere. For observations taken on 2008\,Nov\,03 no suitable calibrator was observed in HIGH-SENS mode (photometry exposures are taken after the fringe exposures) during that night. Therefore, one calibrator observed in SCI-PHOT mode (interferometric and photometric signal are recorded simultaneously) was used to calibrate the data. As the photometry of this calibrator was also recorded independently right after the simultaneous recording of interferometry and photometry, we were able to reduce the calibrator within the same (HIGH-SENS) mode as the science target.\\
Error bars are derived from the standard deviation of the visibility profiles calibrated with the given calibrators. If the derived error was lower than 10\% or only one calibrator was available, a standard multiplicative error of 10\% was assumed\footnote{Except for the observations obtained on 20080\,Nov\,03 (see Sect.\,\ref{quality}).}, as the error introduced by the time difference in the recording of fringes and photometry lies between 7--15\,\% (Chesneau 2007).

\begin{table}[]
\begin{center}
\begin{footnotesize}
\caption{\label{calibrators}Properties of the calibrator stars.}
\begin{tabular}{lllll}
\hline\hline
HD & Name&Sp. T.&$F_{12}$\tablefootmark{c} & $\theta$\\
 &&& [Jy] & [mas]\\
\hline
224935&YY\,Psc & M3\,III&86.90&$7.10\pm0.07$\tablefootmark{a} \\
18884&$\alpha$\,Cet &M1.5\,IIIa&234.70 &$11.02\pm0.04$\tablefootmark{a} \\
12929&Hamal&K2\,III&77.80 & $7.43\pm0.52$\tablefootmark{b}\\
\hline
\end{tabular}
\end{footnotesize}
\end{center}
\tablefoottext{a}{\tt{http://www.eso.org/observing/dfo/quality/MIDI/qc/\\calibrators\_obs.html}}\\
\tablefoottext{b}{\tt{http://www.eso.org/observing/etc/}}
\tablefoottext{c}{\tt{http://simbad.u-strasbg.fr/simbad/}}
\end{table}
\begin{figure}
\centering
\includegraphics[width=8.5cm, angle=180]{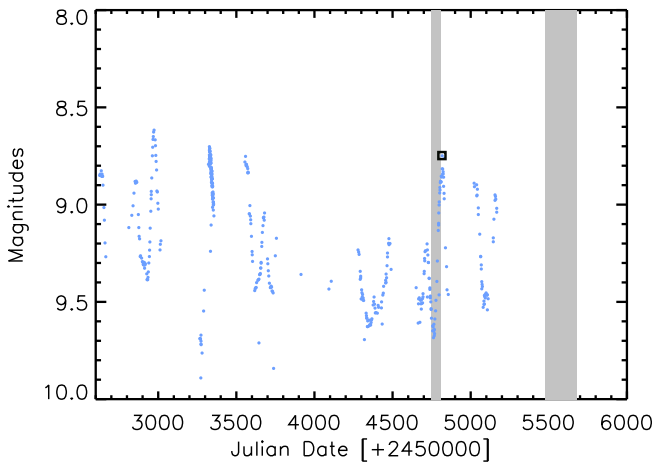}
\caption{\label{asas} ASAS light curve of SV\,Psc. The grey shaded areas mark the range of the MIDI observations. The black square gives the adopted maximum for the phase determination.}
\end{figure}

\subsection{Data quality check}
\label{quality}
During the data reduction the following selection criteria were applied to the science target as well as to the calibrator targets that were used to calibrate the final visibilities, differential phases and errors: (i) the differences in the calibrated visibilities between MIA and EWS were within 5-10\%, (ii) the Fourier amplitude histogram given by MIA was close to a narrow Gaussian distribution, (iii) the raw visibility was smaller than unity, (iv) the power spectral density distribution given by MIA was narrow-peaked and most of the signal was located within the expected frequency range, and (v) the dSky parameter that can be adjusted by MIA (determining the region where the sky background is measured) had the optimal value for the given data set. Observations were excluded from the astrophysical interpretation if one or more of the aforementioned criteria were encountered. 
Data sets flagged with `a' in Table~\ref{journal} were excluded because of poor atmospheric conditions which lead to large systematics in the calibrated measurements (Chesneau 2007). Data sets flagged with `b' were not used in the interpretation because the differences between both reduction softwares MIA and EWS were larger than 10\% (irrespective of the mask used).\\
Observations obtained in 2008 Nov 03 show broader power spectral density distributions (PSD) than the others. This can introduce a systematic bias, leading to an underestimation of the calibrated visibility amplitude. Indeed, Fig. 4 shows that the visibility from 2008 Nov 03 (dark grey line) is lower than that from 2008 Dec 06 (light grey line). We therefore applied some tests to ensure that the change in visibility level is caused by the morphology of the star and not by the quality of the observations. Three pairs of observations of another science target were used. Each pair was observed close in time at similar baseline lengths and position angles with good PSD (i.e. comparable to the 2008 Dec 06 data set) and bad PSD (i.e. comparable to the 2008 Nov 03 data sets). The corresponding calibrated visibilities were then compared to each other. We found a systematic underevaluation of the calibrated visibilities between 5 and 25\% in the case of bad PSD. 
We then tested three approaches: (i) correcting for the visibility amplitude adding up a 25\% amplitude level to the calibrated visibilities; (ii) increasing the uncertainty level by 25\% (see Fig.\,\ref{close-base}) in a way that the error bars on the calibrated visibilities are $\pm$35\% (standard calibration plus systematic bias); (iii) not taking observations of 2008 Nov 03 into account.\\
Approaches (i) and (ii) still show a difference in the visibility amplitude between PA=73$^\circ$ and PA=126$^\circ$ (see Fig.\,\ref{close-base}). The model fitting with GEM-FIND (see Sect.\,\ref{fittingstrat}) was performed for all three approaches. In the case of approach (iii) the degree of freedom is not large enough to allow the use of the two component models (see Sect.\,\ref{fittingstrat}). However, by using one component models we were still able to show that our observations can be reproduced best with elliptical models. Different two component as well as one component models where fitted for approaches (i) and (ii).  The obtained results were very similar. Given this similarity, we decided to take only the second approach into account in the following sections.

\subsection{$N$-band spectrometric variability}
\label{nbandvar}
\begin{figure}
\centering
\includegraphics[width=6.5cm, angle=90]{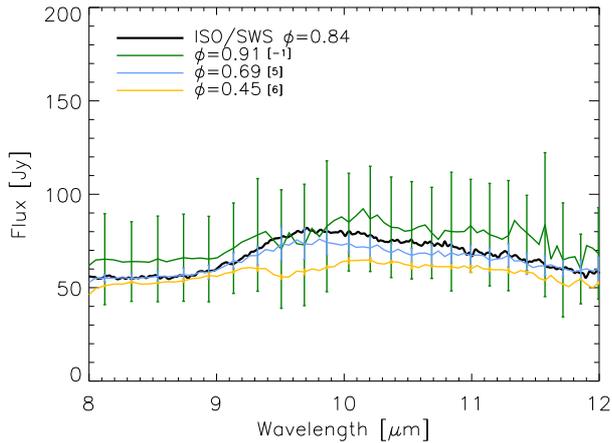}
\caption{\label{iso} Averaged MIDI flux (thin full lines) for different visual phases and overplotted ISO/SWS flux (thick black full line) of SV\,Psc in the $N$-band. Error bars denote the standard deviation of all observed fluxes for the respective night and telescope. The ISO SWS spectrum was observed in 1998 (i.e.\,33 cycles before the first MIDI observations).}
\end{figure}

\begin{figure}
\centering
\includegraphics[width=6.5cm,angle=90]{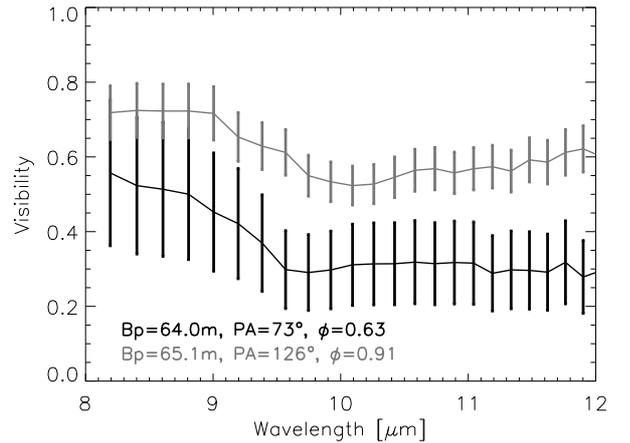}
\caption{\label{close-base} Spectrally dispersed MIDI visibilities of SV\,Psc for observations taken at similar projected baseline lengths but different position angles.}
\end{figure}

In Fig.~\ref{asas} the light curve of SV\,Psc is shown (ASAS, Pojmanski 2002). The grey shaded areas mark the range of the MIDI observations. The visual phase of the MIDI observations is derived assuming a variability period of 116.33\,d (taken from ACVS, Pojmanski 2002). The adopted phase-zero point (corresponding to the visual maximum of the star) for the phase determination is $T_0=2\,454\,817$ and is marked as a black square in Fig.~\ref{asas}.
In Fig.~\ref{iso} the MIDI spectra for different visual phases are overplotted with the ISO/SWS spectrum (Sloan et al.\,2003).  The flux level of the MIDI observations is the same (within the error bars) as the one of ISO/SWS, implying that the level of $N$-band emission has not changed from 1998 to 2011. 



\section{Morphological Interpretation}
\label{models}
Figure\,\ref{close-base} shows the MIDI calibrated visibilities for two selected data sets that were taken at approximately the same baseline length but at different position angles.  
A comparison of these two measurements indicates a clear difference in the visibility level and also in the spectral shape, where the drop in the visibility between 8 and 9.5\,$\mu$m is nearly twice as large for observations at a position angle of 73$^\circ$. As the visibility profile of a spherical source should be similar irrespective of the position angle, the circumstellar environment of SV\,Psc can be assumed to be not spherically symmetric. Moreover, the mid-infrared stellar spectra (see Fig.~\ref{iso}) do not show any significant variation from phase to phase and cycle to cycle. Finally, as reported by Karovicova et al. (2011), no intra-cycle/cycle-to-cycle variations in visibility have been detected for the oxygen-rich Mira variable RR\,Aql which presents a visual lightcurve amplitude six magnitudes larger than the one of SV\,Psc. Therefore we can be confident that the difference in the visibility amplitude presented in Fig.~\ref{close-base} is related to a geometrical effect only.\\
This difference directly proves that the spherically expanding wind models suggested by Knapp (1998) and Kahane \& Jura (1996) are unlikely to be responsible for the double-component CO line profile in the case of SV~Psc.\\
In the following, the morphology of the close circumstellar environment of SV\,Psc will be constrained by fitting visibility measurements with two-dimensional, chromatic geometrical models.


\subsection{Fitting strategy}
\label{fittingstrat}
In order to interpret the MIDI interferometric data, the software GEM-FIND (GEometrical Model Fitting for INterferometric Data) was developed. GEM-FIND is  written in the IDL language and allows fitting the spectrally dispersed visibility measurements with centro-symmetric and asymmetric, chromatic geometrical models. Each model includes a set of wavelength-independent and/or wavelength-dependent parameters. A detailed description of the fitting strategy and $\chi^2$-minimization can be found in Appendix \ref{chiappendix}.\\
\begin{table}[]
\begin{center}
\begin{footnotesize}
\caption{\label{modelparam}Parametric description of the best fitting geometrical models found in this study.}
\begin{tabular}{llllll}
\hline\hline
Model   \rule{0pt}{2.6ex}                         & \multicolumn{2}{l}{$\lambda$ independent} & $\lambda$ dependent& $\chi_{\rm min}^2$\\
& fixed & grid&&\\
\hline
UD+Dirac               &                 & $\Delta\alpha$, $\Delta\delta$      &  $f$, $\theta_{\mathrm{prim}}$ &  0.60    \\
CircUD+EllGauss  & $\theta_{\mathrm{cen}}$                &  $\psi$, $\eta$ &  FWHM$_{\mathrm{maj}}$, $f$        &   0.49   \\
\hline
\end{tabular}
\end{footnotesize}
\end{center}
\tablefoot{$\Delta\alpha$, $\Delta\delta$\ldots angular offset of the binary component from the primary star; $f$\ldots flux ratio binary/primary or central star/envelope; $\theta_{\mathrm{prim}}$\ldots the diameter of the primary component; $\theta_{cen}$\ldots diameter of the central star; $\psi$\ldots inclination angle of the ellipse; $\eta$\ldots axis ratio of the minor/major axis of the ellipse;  FWHM$_{maj}$\ldots Full Width at Half Maximum of the  Gaussian distribution major axis; }
\end{table}
\begin{figure*}[!t]
\centering
\includegraphics*[width=13.cm,bb=81 540 550 707]{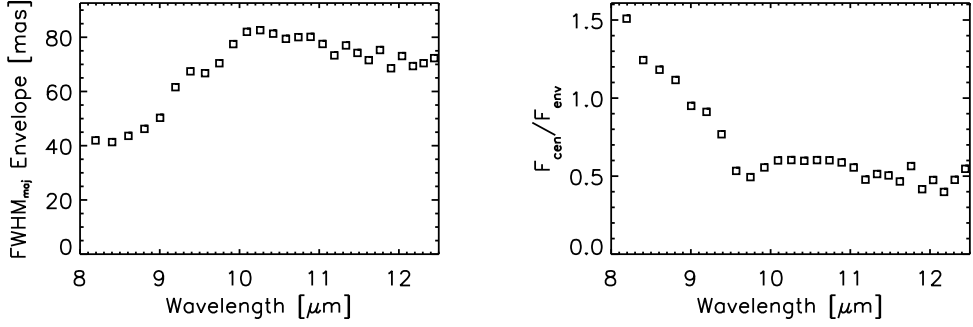}
\caption{\label{diskparam} Wavelength-dependent FWHM of the envelope's major axis and flux ratio of the central star over the envelope determined from the best-fitting disk model.}
\end{figure*}
\begin{figure*}[]
\centering
\includegraphics[width=6.8cm]{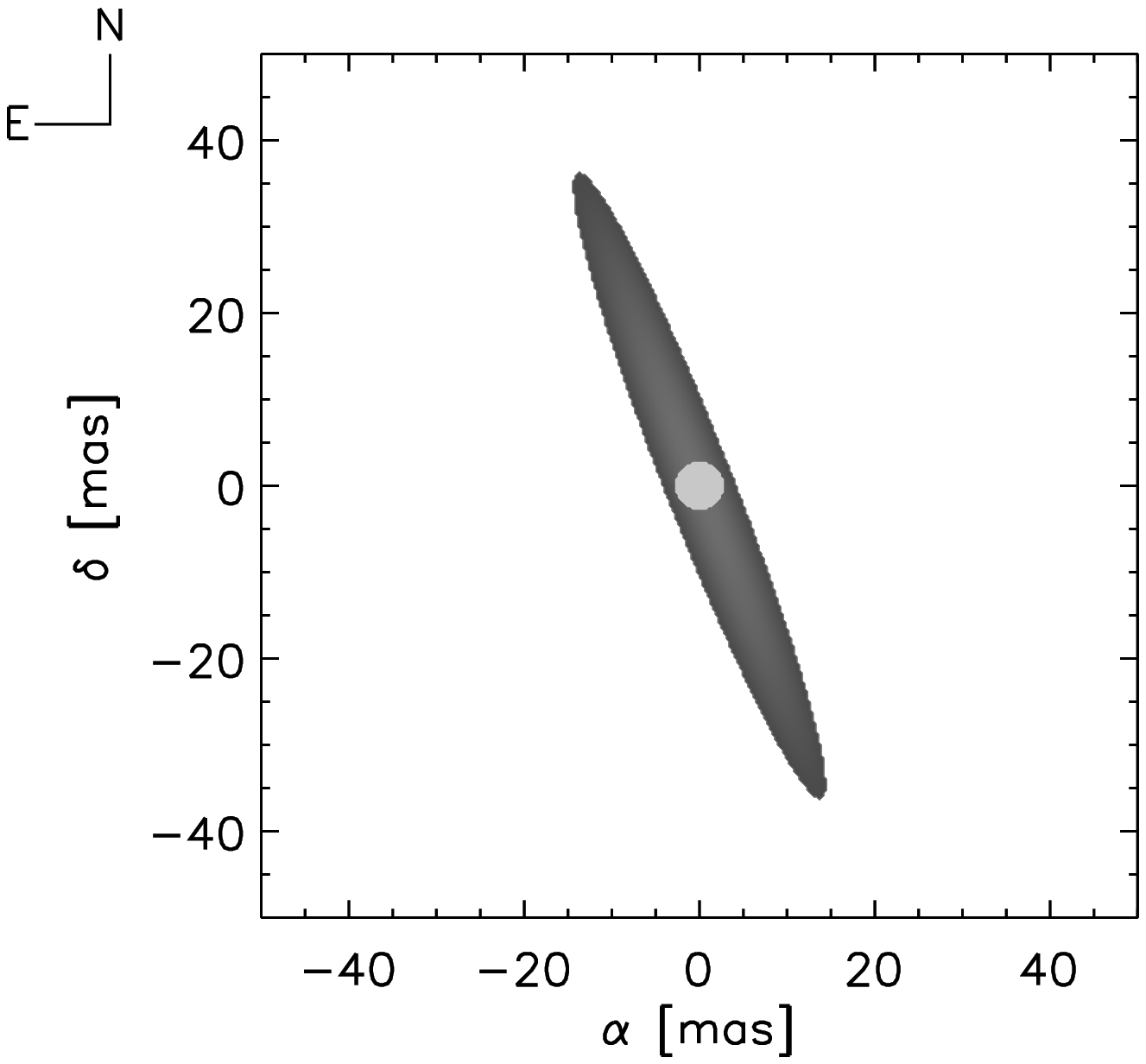}
\includegraphics[width=6.8cm]{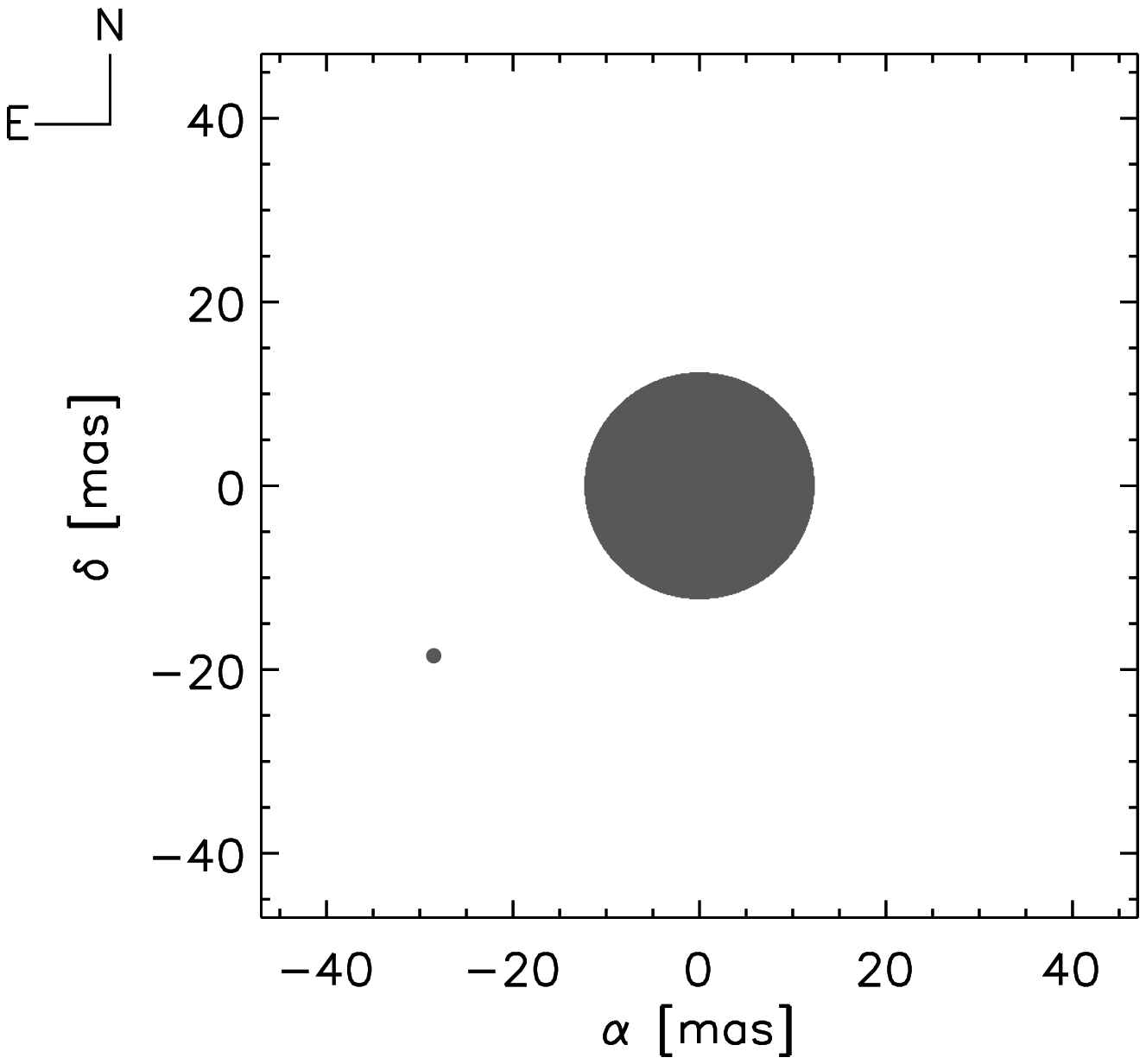}
\caption{\label{skyplane} Schematic view of the best-fitting models (left panel: disk model, right panel: binary model) in the sky plane at 10\,$\mu$m. Note that the binary component is considered to be a point source. Only one of the two possible symmetric solutions for the binary component location is presented here.}
\end{figure*} 
We used different centro-symmetric and non-centro-symmetric models having an optically thick (i.e.\,one-component models) or optically thin (i.e.\,two-component models) circumstellar environment. All spherically symmetric models show a large $\chi_{\mathrm{min}}^2$ implying that those models are not able to reproduce the data. This is consistent with the results of Fig.~\ref{close-base}. The optically thick elliptical models are also unable to provide a good fit to the data. This can be explained by the fact that the circumstellar environment of SV\,Psc is optically thin enough (i.e.\,$\dot{M}=1.2\times10^{-7}$\,$M_{\odot}$\,yr$^{-1}$) to allow the radiation of the central star to escape. An optically thin two-component model consisting of a central star (represented by a Uniform Disk distribution) and an elliptical dust envelope (represented by a Gaussian distribution) allows a much better fit to the data. Additionally, a two-component model consisting of a resolved star (represented by a Uniform Disk distribution) and an unresolved companion (represented by a Dirac delta function) was fitted to the data. The model parameters and the color-reduced minimum $\chi_{\mathrm{min}}^2$ of the best fitting models are given in Table~\ref{modelparam}. In the following, the two best fitting models are described in more detail.


\subsection{Disk model}
\label{disk}
The circumstellar environments around low mass-loss AGB stars are generally considered to be optically thin (e.g. Kemper et al. 2001). We therefore expect  the stellar emission to dominate over the emission of the dust envelope at short mid-infrared wavelenghts (i.e. $\sim$\,8 - 9\,$\mu$m: see Sacuto et al. 2008). A geometrical model consisting of a Uniform Disk and an elliptical Gaussian representing a central star surrounded by a disk, respectively, was used.

\subsubsection{Visibility modeling}
\begin{figure*}[!t]
\centering
\includegraphics[width=14cm,bb=81 421 350 587]{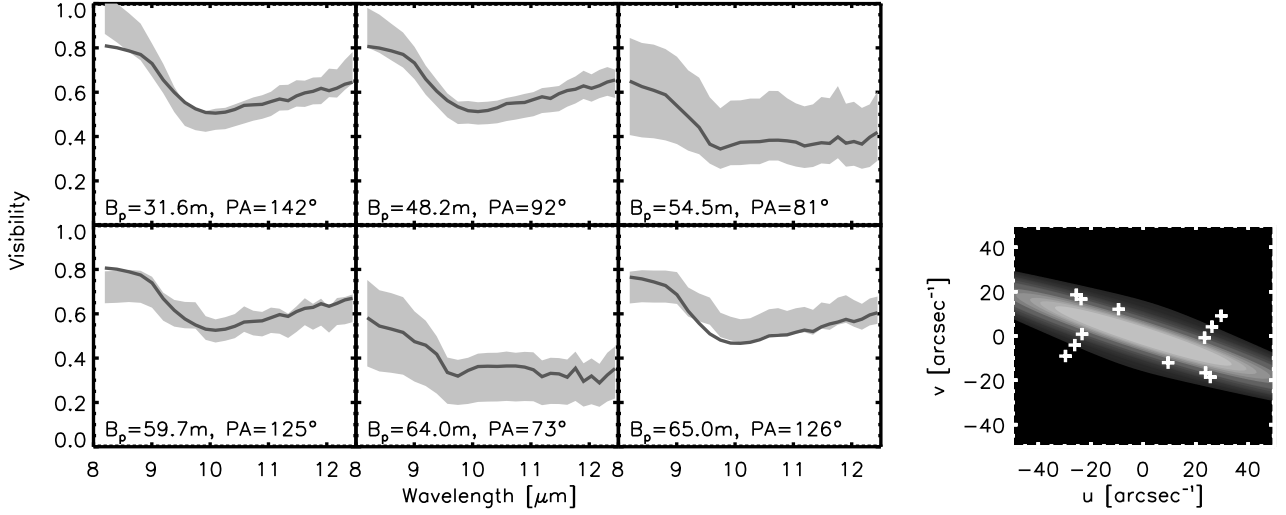}
\caption{\label{diskvisibilities} Best-fitting disk-model visibilities (dark-grey line) superimposed on the calibrated MIDI visibilities (grey-shaded area). Lower right plot: Modeled visibility distribution in the $uv$-plane at 10\,$\mu$m plotted with the $uv$-coverage of the observations.}
\end{figure*}

The analytical expression of the visibility for the disk model can be found in Appendix\,\ref{formulaeappendixdisk}.
This model consists of four free parameters: the Full-Width-at-Half-Maximum of the major axis FWHM$^{\lambda}_{\mathrm{maj}}$, the flux ratio of the central star over the envelope $f_\lambda$,  the ratio of the minor to the major axis of the ellipse $\eta$ and the inclination angle of the ellipse $\psi$. The parameter FWHM$^{\lambda}_{\mathrm{maj}}$ is chosen to be wavelength dependent to take the chromatic variation of the opacity of the dusty environment into account. This results in a variation of the dimension of the structure. The parameter $f_\lambda$ is also chosen to be wavelength dependent considering that the emission at shorter wavelengths (8-9.5\,$\mu$m) results from the warmer photospheric regions, whereas at longer wavelengths (9.5-12.5\,$\mu$m) the emission is dominated by the cooler dusty environment. The parameters $\eta$ and $\psi$ are considered to be wavelength independent. 
The diameter of the central star is fixed to 5.5\,mas (derived from the $V-K$ color index, van Belle et al., 1999)Ó and is therefore not considered as a free parameter. 
The best-fitting model yields a $\chi_{\rm min}^2$ of 0.49 with the major axis of the disk inclined by an angle of 21$^\circ$$^{+9^\circ}_{-6^\circ}$ North-East and the axis ratio being 0.1$^{+0.4}_{-0.0}$ leading to the disk-like structure to be seen almost edge-on. The errors on the wavelength independent parameters are given as the upper and lower limiting values of the confidence interval at a confidence level of 68\%. The wavelength-dependent parameters are plotted in Fig.~\ref{diskparam}. The FWHM of the major axis of the elliptical envelope steeply increases between 8 and $\sim$10\,$\mu$m by almost a factor of 2. Driebe et al.\,(2008) found that this signature is typical for a silicate dust environment. The flux ratio of the central star over the envelope, on the other hand, steeply decreases in this wavelength range. This can be explained by the fact that the emission of the central star becomes less significant after 9.5\,$\mu$m due to the contribution of the silicate-rich, dusty environment.\\
The left plot in Fig.~\ref{skyplane} shows a schematic view of the structure for the best-fitting disk plus envelope model in the sky plane at 10\,$\mu$m. \\
The model visibilities derived with the best-fitting set of parameters are overplotted to the observed ones in Fig.~\ref{diskvisibilities}. 
The lower right plot in Fig.~\ref{diskvisibilities} shows the visibility distribution of the best-fitting model in the Fourier plane at 10\,$\mu$m where the white crosses represent the $uv$-coverage of the observations. 
\begin{figure*}[]
\centering
\includegraphics[width=9.cm,bb=81 421 250 587]{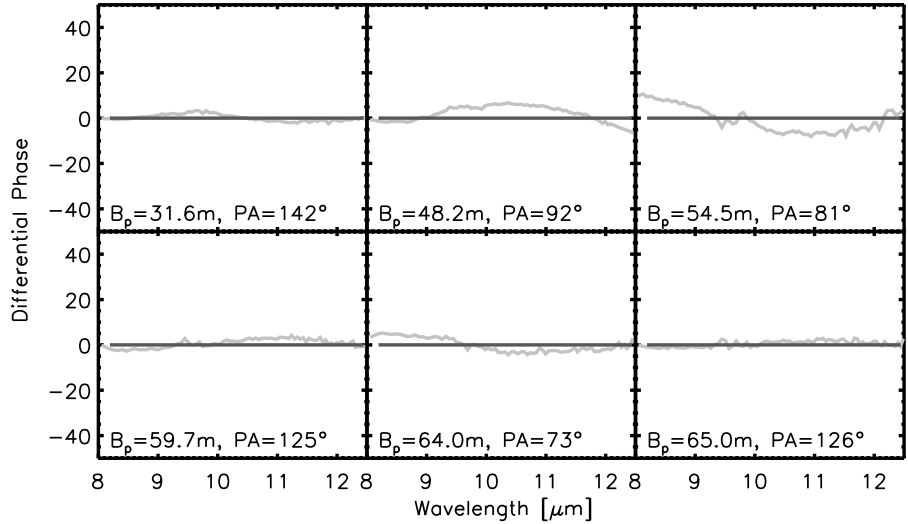}
\caption{\label{diskphases} Best-fitting disk-model differential phases (dark-grey line) superimposed on the calibrated MIDI differential phases (light-grey line). }
\end{figure*}   
\subsubsection{Differential phase constraints}
The observed fringe phase is the visibility phase plus an unknown error that is introduced by atmospherical changes (e.g.\,Haniff 2007). These changes can be considered as achromatic, because the refractive index of air only slowly varies with wavelength, i.e.\,errors can be related between different wavelengths (Jaffe 2004, Tubbs et al.\,2004). The subtraction of a mean phase and gradient as a function of wavenumber therefore removes most of those errors. This so-called differential phase provides information on possible asymmetry of the source within the spatial and spectral region covered by the different interferometric measurements. For example, strong phase signals were detected in sources harboring a circumcompanion disk or a circumbinary structure (Ohnaka et al.\,2008, Deroo et al.\,2007, respectively).
Model phases were derived from the best-fitting model complex visibilities and were computed according to the description by Tubbs et al.\,(2004). They find an accuracy of the observed differential phase of 1$^\circ$ RMS for stable atmospheric conditions. However, changes in humidity during measurements can cause larger errors. For the measurements obtained for SV\,Psc the humidity and seeing levels were quite stable throughout the night. Therefore, an error of $\sim$5$^\circ$ on the measurements of the differential phases seems reasonable (Ohnaka et al. 2008; Deroo et al. 2007).\\
Observed and modeled differential phases are plotted in Fig.~\ref{diskphases}. As the model is centro-symmetric, we do not expect any signature in the differential phase. 
The curvature in the observed differential phase signatures at 54.5\,m and 64.0\,m is caused by the different indices of refraction of air present in the VLTI delay line tunnels (Deroo et al.\,2007). This is normally corrected in the data reduction process by using the instrumental phase. However, in the case of the 54.5\,m baseline the difference in time of observation of the target and the one of the calibrator is large, implying that the instrumental phase is not able to correct for this curvature. On the contrary, the phase signature for the 48.2\,m baseline seems to be real. Such a signature in the differential phase would point to an asymmetrical brightness distribution. A binary companion could account for this asymmetry.


\subsection{Binary model}
\label{binary}
The binary model was constructed considering a geometrical model consisting of a Uniform Disk (representing the resolved primary component) and a Dirac delta function (representing the unresolved secondary component). 
\subsubsection{Visibility modeling}
\label{vis}
The analytical expression for the visibility of the binary model can be found in Appendix\,\ref{formulaeappendixbinary}.
This binary model consists of four parameters: the angular diameter of the primary component $\theta^{\lambda}_{\mathrm{prim}}$, the flux ratio of the binary component over the primary $f_\lambda=\mathrm{F^\lambda_{bin}}/\mathrm{F^\lambda_{prim}}$ and the angular coordinates of the secondary relative to the primary  $\Delta\alpha$ and $\Delta\delta$.
The parameter $\theta^{\lambda}_{\mathrm{prim}}$ represents the central star plus an extended atmosphere and is chosen to be wavelength dependent in order to account for the chromatic variation of the opacity of the environment. This leads to a variation of the dimension of the AGB primary component. The parameter $f_{\lambda}$ is also chosen to be wavelength dependent. This considers that the emission at shorter wavelengths (8--9.5\,$\mu$m) is dominated by the primary star while at longer wavelengths (9.5--12.5\,$\mu$m) the dust material surrounding the primary star and the material accreted by the binary component are emitting stronger. The angular offset  ($\Delta\alpha$ and $\Delta\delta$) of the component, on the other hand, is assumed to be wavelength independent. The fitting strategy is the same as described in Appendix~\ref{chiappendix}. Figure~\ref{chisquare} shows the color-reduced $\chi^2$ distribution for the geometrical model. The joined confidence region of the wavelength independent parameters that results from the best fit to the MIDI data is displayed by the white contour line. 
For our data a confidence level of 68\% was used. Note that two symmetrical solutions exist, because the visibility determination does not allow the determination of the sign of the separation vector (Ratzka 2009). The best-fitting model yields an $x$-position of the binary of $\Delta\alpha=-29^{+9}_{-5}$\,mas, a $y$-position of $\Delta\delta=-18^{+8}_{-12}$\,mas (and for the symmetrical solution: $\Delta\alpha=29^{+5}_{-9}$\,mas, $\Delta\delta=18^{+12}_{-8}$\,mas) with a minimum color-reduced $\chi_{min}^2$ of 0.60. Assuming a distance of 400\,pc this position corresponds to a separation of $13.7^{+4.2}_{-4.8}$\,AU. The position of the binary is well constrained by our model. A schematic view of the best-fitting binary model is plotted in Fig.~\ref{skyplane}. The wavelength-dependent parameters $\theta^\lambda_{\mathrm{prim}}$ and f$_{\lambda}$ are plotted in Fig.~\ref{parameters}. The increase of the angular diameter of the primary component from 8 to 9.5\,$\mu$m can be explained by the larger opacity caused by the silicate dust material present in the atmosphere of O-rich AGB stars. The increase of the flux ratio of the binary component over the primary from 8 to 9.5\,$\mu$m can be attributed to the fact that after 9.5\,$\mu$m the silicate dust emission of the material that is accreted by the binary component is stronger. 
\begin{figure}[!h]
\centering
\includegraphics[width=13.cm]{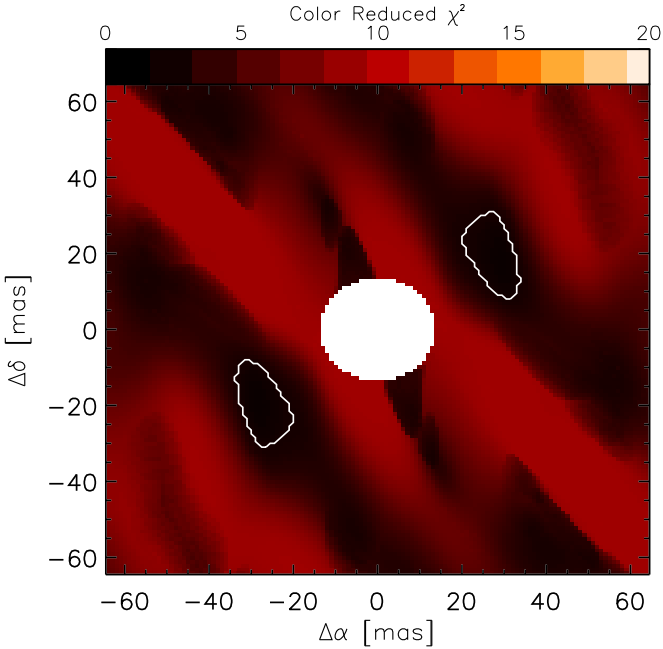}
\caption{\label{chisquare}  Color-reduced $\chi^2$ map for the grid of wavelength-independent parameters of position of the binary component. The white area in the middle marks the region where the primary component is located. In this part, no fitting was performed. White contours are drawn at $\chi^2_{\mathrm{min}}+1.15$ indicating the 68\% confidence level. }
\end{figure}
\begin{figure*}[!t]
\centering
\includegraphics*[width=13.cm,bb=71 540 550 707]{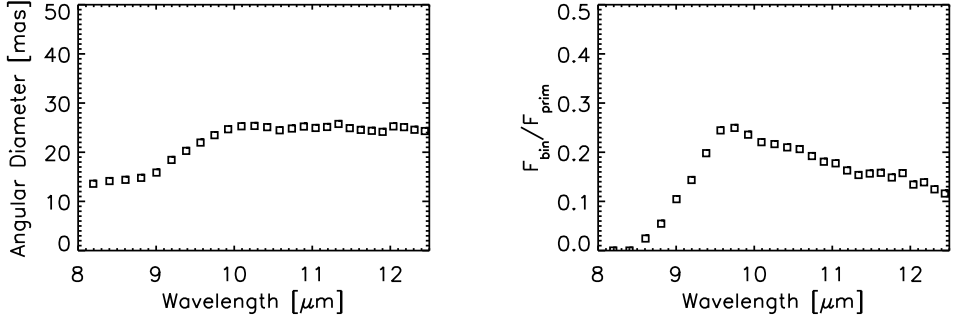}
\caption{\label{parameters} Angular diameter of the primary and flux ratio of the binary component over the primary determined from the best fitting binary-model.}
\end{figure*}
The visibilities obtained from the model with the best-fitting parameters are overplotted on the MIDI visibilities in Fig.~\ref{visibilities}. The lower right plot in Fig.~\ref{visibilities} shows the visibility distribution of the best-fitting model in the $uv$-plane at 10\,$\mu$m. An increase of the $uv$-coverage using baselines perpendicular to the direction of the binary axis would help to ensure the presence of the binary companion.

\subsubsection{Differential phase constraints}
A binary system will be interpreted by an interferometer as an asymmetrical brightness distribution, which means that we expect a differential phase signature. The instrumental and atmospherical corrections that are applied to the measured MIDI phase by the data reduction software EWS involve the removal of a wavenumber-dependent slope and offset from the phase (Jaffe 2004). The differential phase of a binary system will therefore oscillate around zero (see Fig.\,C1 in Ratzka et al.\, 2009 for an example). The missing offset and slope will prevent the observer from getting absolute astrometric information (Millour et al.\,2011), but will still allow to determine the position of the secondary with respect to the primary. The differential phase is also the only way to disentangle between both symmetric solutions of the binary location.
The model differential phases for both symmetric solutions (dark-grey full line corresponding to the upper right solution in Fig.\,\ref{chisquare} and dark-grey dotted line corresponding to the lower left solution) are overplotted on the MIDI differential phases in Fig.~\ref{phases}. There is no full agreement between the model and the differential phase measurements. However, because of the simplicity of the model that is used, we cannot exclude that a more sophisticated model, involving the full circumbinary structure, would smooth the signature revealed in the differential phase resulting in a better agreement with the measurements.\\
The lower right plot in Fig.~\ref{phases} shows the modeled phase distribution and the observations at 10\,$\mu$m in the $uv$-plane. According to this plot larger baseline lengths are needed to measure a significant phase signature as the central part (probed by the $uv$-coverage of the observations) is blurred by the resolved, spherical primary component. 


\section{Discussion}
\label{discussion}
\begin{figure*}[!t]
\centering
\includegraphics[width=14cm,bb=81 421 350 587]{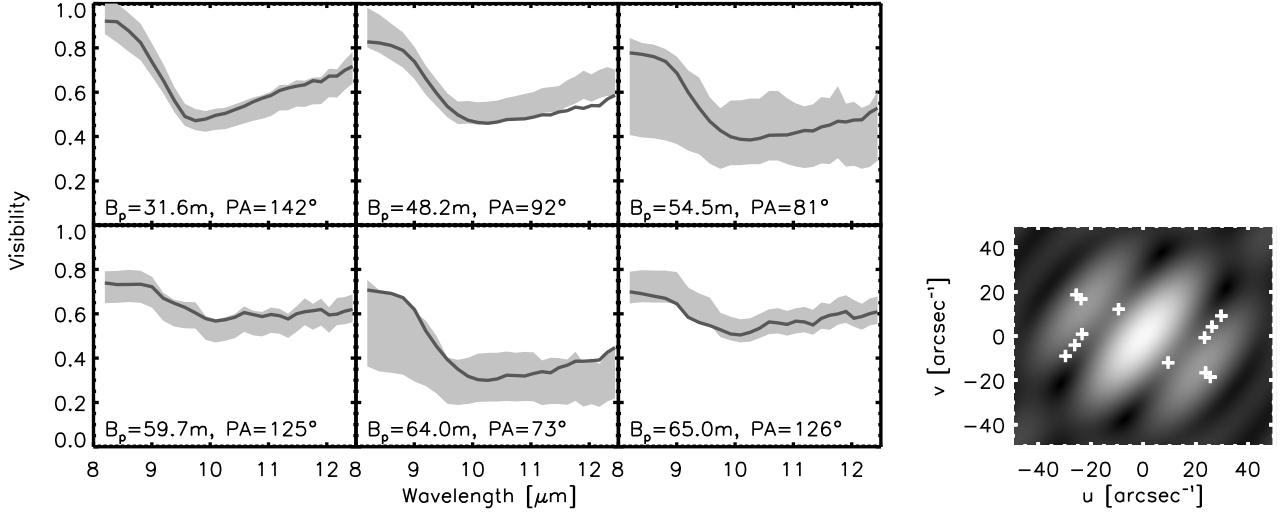}
\caption{\label{visibilities} Best-fitting binary-model visibilities (dark-grey line) superimposed on the calibrated MIDI visibilities (grey-shaded area). Lower right plot: Modeled visibility distribution in the $uv$-plane at 10\,$\mu$m plotted with the $uv$-coverage of the observations.}
\end{figure*} 
\begin{figure*}[!t]
\centering
\includegraphics[width=14cm,bb=81 421 350 587]{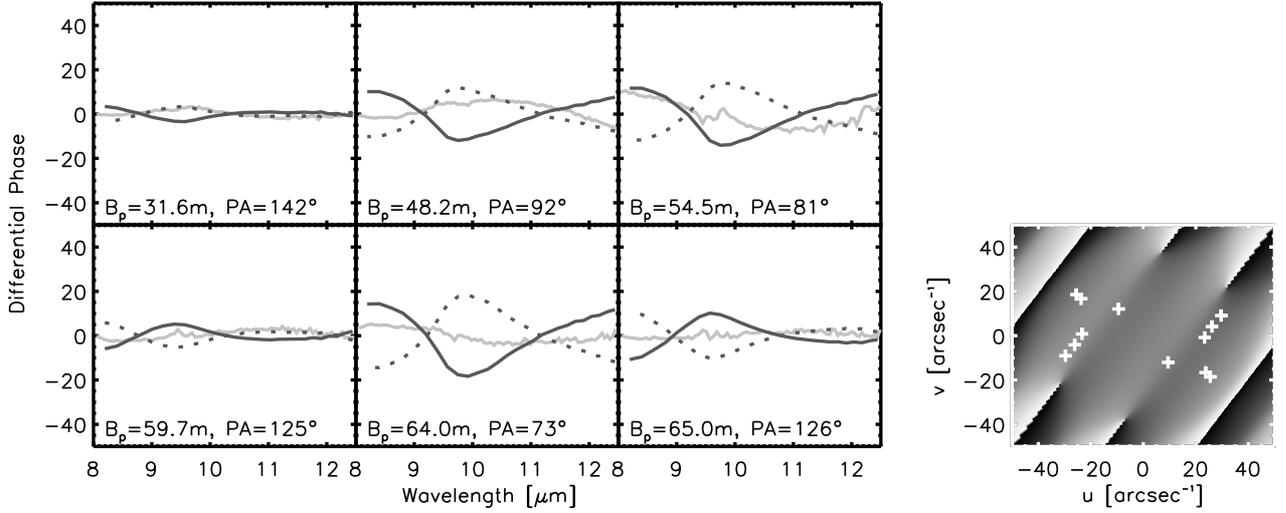}
\caption{\label{phases} Best-fitting binary-model differential phases (dark-grey full and dotted line corresponding to upper right and lower left solution in Fig.\,\ref{chisquare}, respectively) superimposed on the calibrated MIDI differential phases (light-grey line). Lower right plot: Modeled phase distribution in the $uv$-plane at 10\,$\mu$m plotted with the $uv$-coverage of the observations.}
\end{figure*}   
The confidence region on the position of the binary component (see Fig.~\ref{chisquare}) leads to an angle spreading over 39.9$^\circ$. The orbital period of the binary component was calculated by using the angular separation derived from the best fitting binary model (see Sect.~\ref{binary}). Assuming a mass of the primary of $M_{\rm prim}$=1\,$M_\odot$, a mass of the companion ranging from $M_{\rm bin}$=0.5 to 1\,$M_\odot$ (corresponding to the typical mass of a white dwarf or main sequence star) and a distance of the star of 400\,pc, this leads to an orbital period of $P_{\mathrm{orb}}=38.1^{+20.4}_{-22.6}$\,yr. Because our observations spread over 2.17 years, we expect the component to move by a maximum angle of 50.4$^\circ$ on the plane of the sky. This makes the average motion of the binary companion almost consistent with the uncertainty we found on its position. Considering an orbital plane that is inclined by an angle of 32.9$^\circ$ relative to the plane of the sky, the angle of 50.4$^\circ$ derived from the observed orbital period would decrease by projection and lie within the angle of 39.9$^\circ$. \\
Mastrodemos \& Morris (1999) used 3D smoothed particle hydrodynamics to model the influence of wind velocity, secondary mass and binary separation on the circumstellar environment of AGB stars. Considering their models 12 and 13 (Table~1 of the paper) that have parameters close to the ones we derived for SV\,Psc, the wind morphology would be slightly elliptical or bipolar. 
Moreover, an accretion disk around the secondary component of the size ranging from $\sim$0.6 to 1.8\,AU is predicted by the three-dimensional models. This accretion disk would not be resolved with MIDI making the use of the Dirac distribution to model the binary component reasonable.
On larger scales (few 100 AU) gas and dust that is lost by the star is captured within the binary system and would form a disk-like structure. This structure would develop from the enhancement of the density in the orbital plane. Such a disk could be the reason for the broad feature visible in the CO line profile. In this case a disk in Keplerian rotation, as suggested by Bergmann et al.\,(2000) for RV\,Boo, would be a direct consequence of the presence of a binary companion. Therefore, the solution involving a disk structure seen edge-on does not exclude the presence of a close binary component. In this case MIDI would probe the most inner part (4--16\,AU) of the rotating disk. \\
In order to fully discriminate between disk-like and binary model, a larger $uv$-coverage is necessary. Additional larger baselines are needed to detect a significant signal in the differential phase that would confirm the presence of the binary companion. Position angles perpendicular to the binary axis would allow further constraining of the position of the companion, or constraining the inclination angle of the disk. Moreover, additional measurements would allow  to test more sophisticated models involving a circumbinary structure composed of the AGB star, the close binary component and a disk-like structure.

\section{Conclusion}
\label{conclusion}
In this work we have presented the first high-angular-resolution mid-IR interferometric study of the close circumstellar environment of SV\,Psc using VLTI/MIDI. This star is known to have a peculiar CO line profile that points to a non-spherical circumstellar environment. By comparing observations taken at the same baseline length but at different position angles we found direct proof that the close environment of the object deviates from sphericity. However, the visibility level of two data sets may be affected by some systematic bias. Even if we derived an upper limit for this bias and increased the error bars on these observations accordingly, further observations are needed to fully confirm the results.\\
We developed the software tool GEM-FIND allowing the fitting of centro-symmetric and asymmetric, chromatic geometrical models to the spectrally-dispersed interferometric data. We found two models that are able to reproduce the observed visibilities: (i) a central star surrounded by a highly inclined elliptical envelope having a Gaussian brightness distribution, (ii) a system composed of a resolved primary star and an unresolved binary companion located at a separation of 14\,AU. We find that the wavelength dependent parameters in both models are strongly affected by the silicate-rich, dusty environment.\\
Sensitive UV and X-ray observations could give further constraints on the origin and evolutionary stage of the companion. Futhermore, 
MIDI observations at larger baselines and position angles perpendicular to the binary axis are needed to fully discriminate between these two scenarios. A larger amount of observations would also allow to test more complex models. For example, we cannot exclude the possibility that a binary companion could be the origin of a disk-like density distribution along the orbital plane. This has been shown by theoretical simulations (Mastrodemos \& Morris 1999) and could be tested as soon as additional observations are available. Such a disk could be the reason of the broad feature that is visible in the CO line profile of the star.

\begin{acknowledgements}
This work is supported by the Austrian Science Fund FWF under project number AP23006 and AP23586. The research leading to these results has received funding from the European Community's Seventh Framework Programme under Grant Agreement 226604. This research has made use of the SIMBAD database, operated at CDS, Strasbourg, France. 
\end{acknowledgements}

\appendix

\section{Detailed description of GEM-FIND}
\label{chiappendix}
In order to prevent solutions from falling into local minima, the wavelength-independent parameters included in GEM-FIND are varied using an equally distributed grid.
This applies to all wavelength-independent parameters, except the diameter of the central star, $\theta_{\mathrm{cen}}$, which was fixed 
for the disk model. For each grid point (i.e. combination of wavelength-independent parameters) a non-linear least squares fitting minimization (based on the Levenberg-Marquardt method: see Markwardt\,2008) is performed. This allows determining the best-fitting wavelength-dependent parameters and a color-reduced $\chi^2$ expressed as
\begin{equation}
\chi_i^2=\frac{\sum\limits^j\chi_{j,i}^2}{N} \qquad\mbox{for}\;i=1\dots n, \;j=1\dots N,
\end{equation} 

where $i$ corresponds to the index of a given grid point (i.e. combination of wavelength-independent parameters) and $n$ is the total number of grid points. The corresponding wavelength index is expressed by $j$, while $N$ is the number of wavelength points. The number of degrees of freedom are defined as the number of observations minus the number of free parameters of the model. The best-fitting wavelength-dependent and independent parameters are then determined by the minimum color-reduced $\chi_{\rm min}^2$. Finally, the standard uncertainties of each wavelength-independent parameter are derived from the minimum and maximum values contained in the 68.3\% confidence region defined by
\begin{equation}
\Delta\chi^{2} = \chi_i^{2} - \chi_{\rm min}^{2} = A
\end{equation}
where $A$ depends on the number of wavelength-independent parameters.

\section{Analytical expressions for the visibility}
\label{formulaeappendix}
\subsection{Disk model}
\label{formulaeappendixdisk}
The normalized visibilities of the two components can be expressed as
\begin{equation} 
V_{\mathrm{cen}}=\frac{2J_1(\pi \theta_{\mathrm{cen}} r)}{\pi \theta_{\mathrm{cen}} r} ~~\mbox{ and }~~
\end{equation}
\begin{equation}
V^{\lambda}_{\mathrm{env}}=\exp\left(\frac{-\left(\pi r_{\psi,\eta} \mathrm{FWHM}^{\lambda}_{\mathrm{maj}}\right)^2}{4\ln 2}\right)\;,
\end{equation}
The spatial frequency vector is defined as $r=\sqrt{u^2+\upsilon^2}$. The elliptical envelope is constructed by applying a rotation and compression to one axis that becomes the semi-minor axis. The corresponding spatial frequency vector can be written as
\begin{equation}
r_{\psi,\eta}=\sqrt{u_\psi^2\eta^2+\upsilon_\psi^2} \;,
\end{equation}
where $\eta=a/b$ is the minor to major axis ratio of the elliptical envelope and $\psi$ is the inclination angle of the major axis counted from North to East. The inclined spatial frequencies can be written as 
\begin{equation}
u_\psi=u\cos\psi - \upsilon\sin\psi ~~\mbox{and}~~\upsilon_\psi=u\sin\psi + \upsilon\cos\psi\;.
\end{equation}
The final normalized visibility of the composite model can be written as
\begin{eqnarray}
V(u,\upsilon)&=&\frac{\left|\mathrm{\mathrm{f}_\lambda V_{\mathrm{cen}}}(u,\upsilon)+V^{\lambda}_{\mathrm{env}}(u,\upsilon)\right|}{\mathrm{f_\lambda}+1}\;.
\end{eqnarray}

\subsection{Binary model}
\label{formulaeappendixbinary}
The corresponding normalized visibility of the binary model can be written as 
\begin{eqnarray}
V(u,\upsilon)&=&\frac{\left|\mathrm{V^{\lambda}_{\mathrm{prim}}}(u,\upsilon)+\mathrm{f}_\lambda\exp\left(-2\pi \mathrm{i}(\Delta\alpha u+\Delta\delta \upsilon)\right)\right|}{1+\mathrm{f_\lambda}}\;,
\end{eqnarray}
where
\begin{equation}
\mathrm{V^{\lambda}_{\mathrm{prim}}}=\frac{2J_1(\pi \theta^{\lambda}_{\mathrm{prim}} r)}{\pi \theta^{\lambda}_{\mathrm{prim}} r}
\end{equation}
is the normalized visibility of the primary.

\end{document}